# Two Different Methods for Modelling the Likely Upper Economic Limit of the Future United Kingdom Wind Fleet

Anthony D Stephens and David R Walwyn
Correspondence to tonystephensgigg@gmail.com

## Abstract

Methods for predicting the likely upper economic limit for the wind fleet in the United Kingdom should be simple to use whilst being able to cope with evolving technologies, costs and grid management strategies. This paper present two such models, both of which use data on historical wind patterns but apply different approaches to estimating the extent of wind shedding as a function of the size of the wind fleet. It is clear from the models that as the wind fleet increases in size, wind shedding will progressively increase, and as a result the overall economic efficiency of the wind fleet will be reduced. The models provide almost identical predictions of the efficiency loss and suggest that the future upper economic limit of the wind fleet will be mainly determined by the wind fleet's Headroom, a concept described in some detail in the paper. The results, which should have general applicability, are presented in graphical form, and should obviate the need for further modelling using the primary data. The paper also discusses the effectiveness of the wind fleet in decarbonising the grid, and the growing competition between wind and solar fleets as sources of electrical energy for the United Kingdom.

## Table of Contents

**Table of Figures**

**Table of Tables**

## 1. Managing Wind Surpluses

When wind generation was first introduced to the United Kingdom (UK) National Grid, many in the industry predicted that wind power might one day provide all the grid's needs. Wind variability and intermittency, they argued, could be ameliorated by a combination of energy storage and exchanging surpluses/deficits with countries experiencing different weather patterns. In this paper, we argue that such a strategy would be uneconomic and that there is a likely upper economic limit, the value of which can be estimated using models based on historical wind patterns.

Before discussing the development of the models, we must first consider in more detail the limitations of storage and/or inter-country transfers within the context of the UK wind fleet and the intermittent nature of UK wind generation. As shown in Figure 1, the weekly average wind generation records for 2013 to 2016 reveal a high degree of wind variability from week to week. Apart from wind generation being higher in the 1$^{st}$ and 4$^{th}$ quarters of each year than in the summer, which is to be expected, there is little discernible pattern to the generation.

**Figure 1. Weekly average wind generation; 2013 to 2016**

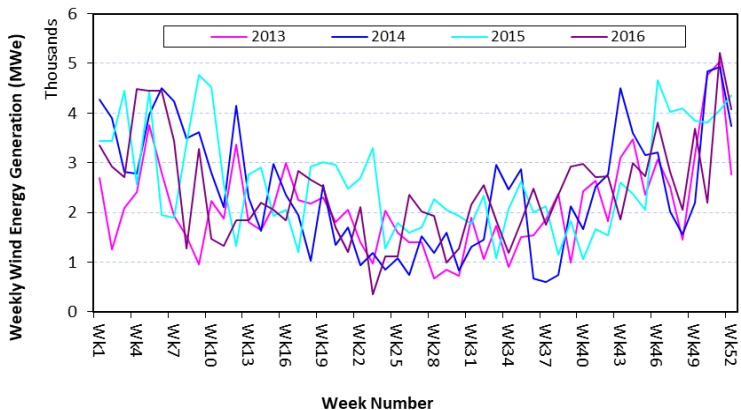

Source: Gridwatch

Furthermore, these weekly averages understate the wind variability; to provide an insight into the impact of real time wind variability on the operation of the National Grid we need to consider real time wind generation records such as those for week 38 of 2016, as shown in Figure 2.

Although the average generation during week 38 of 2016 was 2.36 GW$_e$[1], close to the average for the year, wind generation varied during the week by a factor of 21 to 1, between a low of 0.29 GW$_e$ and a high of 6.20 GW$_e$. Although all the energy from the wind fleet may have been accommodated by the grid during week 38 of 2016, this would not have been the case had the installed capacity of the wind fleet been in the range 34.4 GW$_c$ to 75.3 GW$_c$, the upper limit of the wind fleet suggested by a number of consultants working for the Department of Energy and Climate Change (Royal Academy of Engineering, 2014). The question arises as to whether generation from future larger wind fleets, whose capacity is in excess of demand, might be beneficially used, or must be curtailed.

---

[1] Throughout this paper the suffix e will be used to indicate wind generation e.g. GW$_e$ and MW$_e$, and c to indicate wind fleet capacity e.g. GW$_c$. Other abbreviations, together with a number of terms which have special or restricted meanings are listed at the end of the paper.

**Figure 2. UK wind generation during week 38 of 2016**

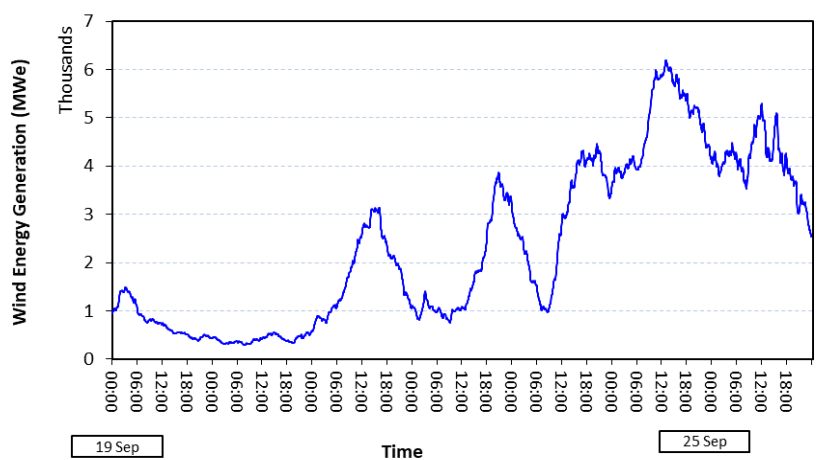

Source: Gridwatch

It is sometimes suggested that electrical storage arrays could be used to store excess wind generation, the stored energy being later returned to the grid. The problem with this suggestion however is that electrical storage arrays are costly, and wind intermittency would lead to a low utilisation of any such systems. For every excess $GW_e$ of generation lasting for 24 hours, 24 $GW_e$h of storage capacity would be needed, equivalent to 186 arrays of the size of Tesla's Lithium ion storage facility in Southern Australia (Edwards, 2017). Although this is currently the world's largest storage array, it has capacity to accommodate only 0.13 $GW_e$h of energy. In 2017, the capital cost of a large Li-ion storage array, including installation and equipment costs, was around $500M per $GW_e$h and. Even if this were reduced to an anticipate $350 million per $GW_e$h by 2024 (Eller and Gauntlett, 2017), a 24 $GW_e$h facility would cost $8.4 billion, equivalent to $43,000 per MWh assuming a 10 year lifetime for the arrays and interest rates of 4% per annum. As we shall see later, occasional excesses of 40 $GW_e$ are to be expected should the wind fleet increase in size to 75$GW_e$, and this would require 960 $GW_e$h of energy storage, equivalent to over 7,000 Tesla storage arrays and a capital requirement of $336 billion. Assuming the stored energy was valued at $40/MWh, 24 $GW_e$h of stored energy would have a value of just under $1 million (vs. a cost of nearly $1 billion).

Storage system only generate income when operational, and the historical data, such as that shown in Figure 2, suggest that electrical storage arrays devoted to handling excess wind generation are unlikely to be in operation sufficiently frequently to make them an economic proposition. Had the wind fleet been sufficiently large to generate 1$GW_e$ in excess of demand on 24[th] September 2016, a 24 $GW_e$h array would have been full to capacity at the end of the day. The array would not then be able either to charge or discharge on the following day when wind generation was still high. Earlier in the week the array would have been inactive because of lack of wind, and it is possible that the array would only have been operational one day that week. Figure 1 reveals many weeks during the years 2013 to 2016 when the array would have been non- operational through lack of wind, including several 3-week periods such as weeks 28 to 30 of 2013, weeks 24 to 26 and weeks 36 to 39 of 2014, and weeks 23 to 25 of 2016. It is clear that the load factor of an array storing excess UK wind generation would be very low, and a combination of high capital cost/low utilisation would make storage of excess generation highly uneconomical. This is consistent with previous studies which concluded that, while utility scale electrical storage might be economic for solar generation, this is not the case for wind generation (Barnhart et al., 2013; Marcacci, 2013).

Inter-country connectors are used extensively to transfer renewable energy surpluses from one country to another, Germany exporting much of its surplus solar generation (Stephens and Walwyn, 2017). We must therefore consider whether excess UK wind generation could be beneficially exported for use in neighbouring countries. As for energy storage, the problem with the use of inter-country transfers for wind power is the intermittent nature of UK wind surpluses. Germany is able to export solar energy on many days a year, resulting in its inter-connectors running at a reasonably high capacity. The unpredictability and variability of UK wind however would result in any interconnectors devoted to exporting UK wind surpluses to its neighbours being used only infrequently. The 2014 Royal Academy of Engineering study concluded that wind speeds in the UK and its continental neighbours are reasonably well correlated (Royal Academy of Engineering, 2014), implying that on those occasions when the UK has wind generation in excess of UK demand, it cannot be assumed that its neighbours would be in a position to accept the excess. Indeed, when the UK was experiencing high winds on 24$^{th}$/25$^{th}$ September 2016 (see Figure 2), Germany was also generating in excess of demand and exporting heavily to its neighbours; there would have been no market in Europe for a UK wind surplus on these days.

The important conclusion is reached that while energy storage and inter-country transfers may in future play a beneficial role in the operation of the UK solar fleet, this is most unlikely to be the case for the UK wind fleet. In the models to be described in the following sections, it will be assumed that whenever UK wind generation is in excess of grid demand the excess generation will be curbed. As the wind fleet increases in size, progressively more generation which is surplus to demand will be curtailed leading to an overall reduction in efficiency of the wind fleet. It is this reduction in efficiency which ultimately determines the upper economic size of the wind fleet.

**2. Simplified Representation of the Wind Fleet**

Although the UK electricity generating system is extremely complex and diverse, when investigating the impact of the wind fleet on the performance of the overall system it is appropriate to make a some simplifying assumptions. A number of generating sources do not need to be considered individually by the model, only in totality, and we shall regard these as a single composite source which we shall call base generation, the latter including the sources of nuclear, biomass energy and imports.

In his study of Irish wind generation Mackay identified dramatic reductions in wind generation with the potential to destabilise the Irish grid (MacKay, 2009). As may be seen in Figure 5, such a rapid reduction in wind generation occurred in the UK on 3$^{rd}$ Nov 2014. On that occasion the rapid reduction in wind generation was compensated for by the rare deployment of Open Circuit Gas Turbine (OCGT) generation. A more logical solution for future larger wind lulls from larger wind fleets would be to run CCGTs continuously on part load, rather than OCGTs. CCGTs are more thermally efficient than OCGTs, are able to run at close to maximum efficiency on part load and can run up to full power from part load in only a few minutes. The part load CCGT generation needed to mitigate possible wind lulls, perhaps 5GW$_e$, must therefore be considered a component of base generation. During the period studied in this paper, 2013 to 2016, base generation was in the range 12 to 15 GW$_e$ but may be significantly different in future decades. Base generation must therefore be treated as an input variable in the models.

A useful means of visualising the contribution of wind generation to grid demand is shown in Figure 3, where wind generation is shown stacked above base generation. The area between grid demand and base generation was traditionally served by coal and gas generation, but in recent years wind generation has been given preferential access to the grid and has displaced coal and gas generation when available. The maximum wind generation which the grid is able to accept at any time is represented by the distance between grid demand and base generation, and we shall call this the wind fleet's Headroom. As may be seen in Figure 3, the wind fleet's Headroom is significantly lower at night than during the day.

**Figure 3. Grid demand and wind generation during week 45 of 2014 (November 2-8)**

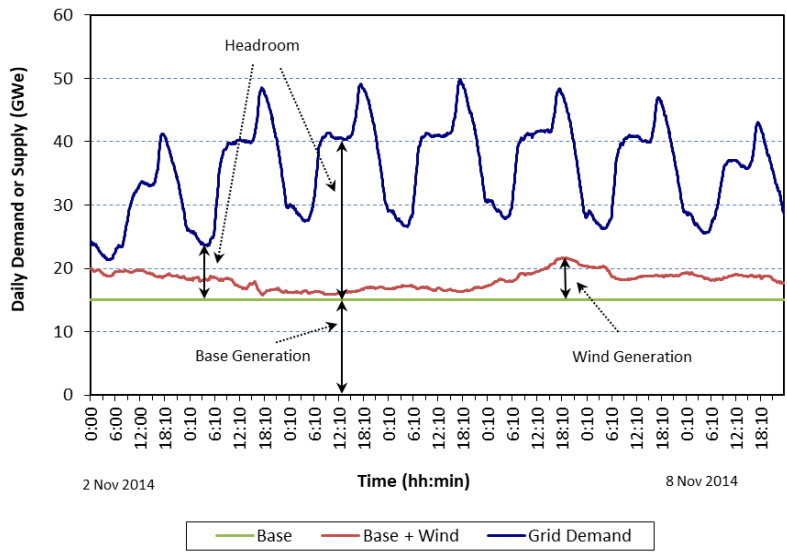

Source: Gridwatch

Grid demand, which has been falling steadily in recent years, should also be considered a model variable. Fortunately, it is possible to define a new variable, annual average Headroom, for which we shall use the symbol Hdrm, as a composite input variable which takes account of changes in both base generation and (annual average) grid demand, as shown in Figure 4, where:

$$\text{Hdrm} = \text{(average annual) grid demand} - \text{base generation} \quad \ldots\ldots \quad \textit{Equation 1}$$

The justification for using the single variable Hdrm to represent both grid demand and base generation was the empirical finding by the authors that replacing real time grid demand records with average weekly grid demand had practically no effect on model predictions (Stephens and Walwyn, 2016). As we shall show later, the model predictions are also insensitive as to whether real time grid demand records or annual average grid demand records are used as model inputs. Over a year, the model predictions of the amount of wind generation a large wind fleet would have to shed through over-production is almost insensitive as to whether grid demand was variable as in Figure 3, or constant at 34.34 GW throughout the year. An explanation for these empirical findings is that the above and below average levels of grid demand in both Figure 3 and Figure 4 compensate for one another.

**Figure 4. Weekly average grid demand during 2014**

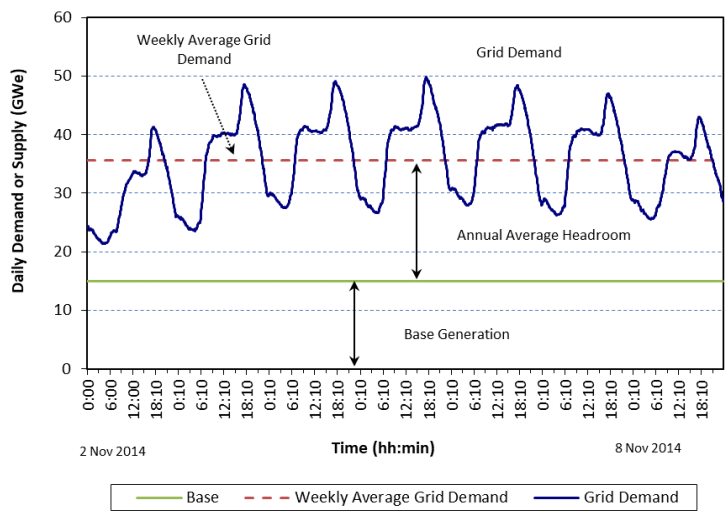

Source: Gridwatch

In the next section, we shall explain how judicious choice of the model input base generation allows predictions of the performance of the wind fleet for different levels of Hdrm, obviating the need to have grid demand as a separate model input. As a result of the relationship between Hdrm, grid demand and base generation, as expressed in Equation 1, results for different levels of Hdrm should embrace all values of base generation and grid demand likely to be encountered in decades to come.

**3.    Model 1: Scaling Real Time Records**

The model, which is in spreadsheet format, takes as input real time Gridwatch records from the web, such as for week 45 of 2014 shown in Figure 3 (Gridwatch, 2017). The model scales these wind generation records to predict what wind generation would have been that week for a range of wind fleet capacities ranging from 10 $GW_c$ to 80 $GW_c$ in steps of 10 $GW_c$, as shown Figure 5.

**Figure 5. Model predictions of wind generation during week 45 of 2014 had the wind fleet capacity been 10$GW_c$ to 80$GW_c$**

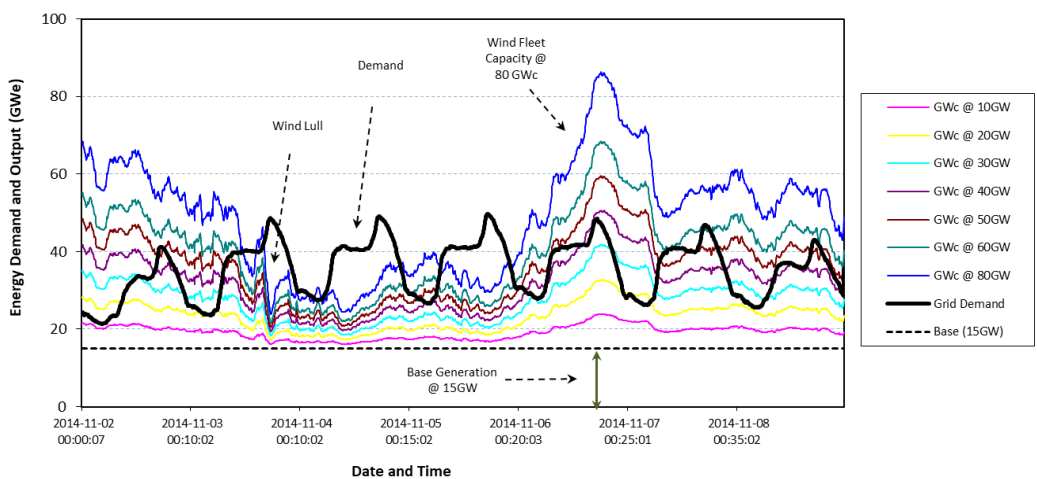

In order to calculate the scaling factors, the model requires as an input value the wind fleet capacity which gave rise to the wind generation recorded by Gridwatch. Unfortunately, the latter only records about 60% of the total UK wind generation. As a consequence, the average wind generation for the year recorded by Gridwatch is divided by the UK government's wind fleet load records. The average generation in 2014 was 2.44 $GW_e$ and the average load factor for the four years 2013-2016 was 0.3075. (Load factors do vary by roughly +/- 10% from year to year but using the average of 0.3075 for the four years puts the results on the same basis and, as will be seen later, introduces little error). The wind fleet wind capacity seen by Gridwatch in 2014 was therefore 2.44/0.3075=7.93 $GW_c$ and is the reference level used to calculate the scaling factors for other sizes of wind fleet. Thus, the scaling factor for a wind fleet of 10$GW_c$ is 10/7.93 (=1.261), for 20$GW_c$ it is 20/7.93 (=2.522) and for 80 $GW_c$ it is 80/7.93 (=10.882). The result of applying these scaling factors to the Gridwatch records of week 45 of 2014 may be seen in Figure 5.

As discussed earlier, wind generation must not exceed grid demand and so, for each time period, the model checks whether the wind generation predictions exceed grid demand and resets it at grid demand if predicted generation exceeds this value. This process is repeated for each week leading to 52 separate weekly spreadsheet models in which wind generation has been capped so that grid demand is not exceeded. Capped predictions, which may be seen in the authors' earlier paper (Stephens and Walwyn, 2016), are then averaged in a separate spreadsheet to produce annual average wind generation, $GW_e$, for different wind fleet capacities, from 10 $GW_c$ to 80 $GW_c$ in steps of 10 $GW_c$. The $GW_e$ predictions for different levels of $GW_c$ allow $GW_e$ vs $GW_c$ curves for different levels of Hdrm to be created, as shown in Figure 6.

Although the model input variable is base generation, the objective was to produce results for different levels of Hdrm rather than base generation and Equation 1 enables us to achieve this objective. Table 1 summarises the values of annual average grid demand during the 4 years 2013 to 2016, and the values of base generation which must be used as model inputs to produce $GW_e$ vs $GW_c$ curves for Hdrm values of 15 $GW_e$, 20 $GW_e$, 25 $GW_e$ and 30 $GW_e$

**Table 1. Choice of base generation values as model inputs to generate the results for Hdrm's of 15 $GW_e$, 20 $GW_e$, 25 $GW_e$ and 30 $GW_e$**

| Year | 2013 | 2014 | 2015 | 2016 |
|---|---|---|---|---|
| Annual Average grid demand ($GW_e$) | 36.10 | 34.34 | 33.075 | 32.35 |
| Base generation to give Hdrm = 15 $GW_e$ | 21.10 | 19.34 | 18.075 | 17.35 |
| Base generation to give Hdrm = 20 $GW_e$ | 16.10 | 14.34 | 13.075 | 12.35 |
| Base generation to give Hdrm = 25 $GW_e$ | 11.10 | 9.34 | 8.075 | 7.35 |
| Base generation to give Hdrm = 30 $GW_e$ | 6.10 | 4.34 | 3.075 | 2.35 |

Figure 6 shows the model predictions for $GW_e$ vs $GW_c$ using the base generation values of Table 1 as inputs so as to generate the curves for Hrdm's of 15 $GW_e$, 20 $GW_e$, 25 $GW_e$ and 30 $GW_e$ for the years 2013 to 2016 (the small spread in the curves between the different years makes it impractical to differentiate individual years in Figure 6).

**Figure 6. $GW_e$ vs $GW_c$ predictions for models using records from the years 2013-2016 and Hdrm's of 15 $GW_e$, 20 $GW_e$, 25 $GW_e$ and 30 $GW_e$**

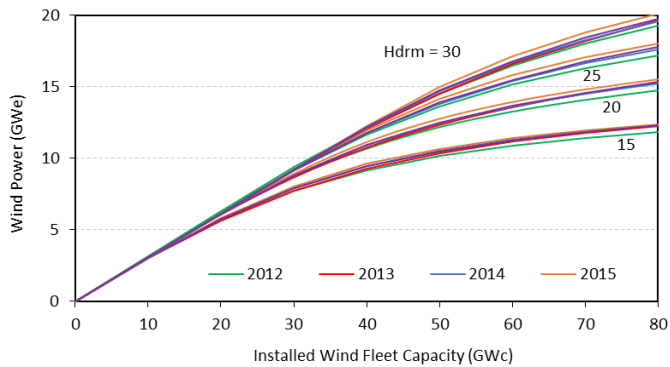

Given the high degree of UK wind variability, such as that seen in Figure 2, it might seem surprising that there is so little variation in $GW_e$ vs $GW_c$ predictions produced using records for the four years. An explanation for this important finding is that, despite this variability, there are only small little differences in annual wind generation records when they are analysed statistically. This may be seen in the wind generation histograms of Figure 7, which shows the proportion of time the wind fleet spent in generation bands of 0.5 $GW_e$ during the years 2013 to 2016. The important implication of the small spread in the curves of Figure 6 is that only a single year's records is needed to generate $GW_e$ vs $GW_c$ curves which have general applicability.

**Figure 7. UK wind generation histograms in bands of 0.5$GW_e$ for 2013 to 2016**

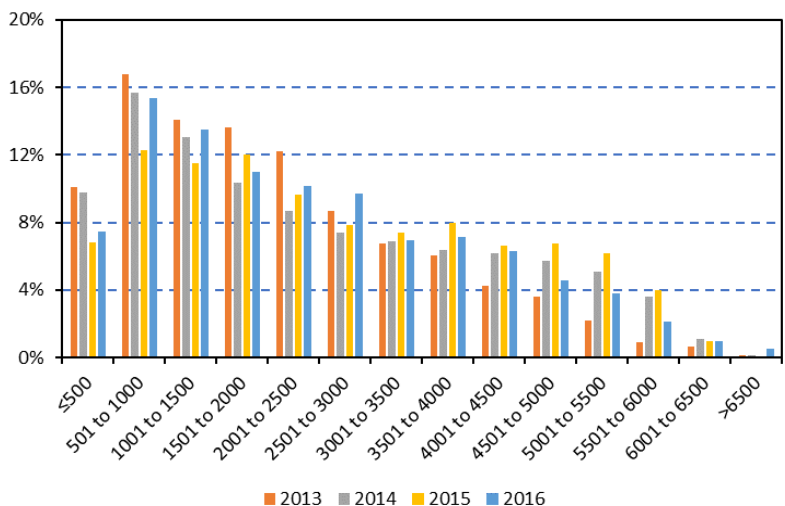

Having developed a model to generate $GW_e$ vs $GW_c$ curves, it is now possible to confirm the suggestion made earlier that such curves are little affected by replacing the real time grid demand records downloaded from Gridwatch with annual average grid demand values. Figure 8 shows, using 2014 data, model predictions of $GW_e$ vs $GW_c$ curves for different levels of Hdrm using real time data (curves without symbols) and a model in which the real time grid demand records were replaced by a constant value of 34.34 $GW_e$ throughout the year (curves with square symbols). An explanation for the two sets of curves of Figure 8 being so similar despite such radically different inputs is that the model which assumes a constant grid demand will over estimate wind shedding in the early and late months of the year, but under-estimate shedding in the summer months (see Figure 4). Figure 8 confirms that the over- and under-estimates largely compensate for each other, particularly for Hdrm values of 20 to 25 $GW_e$, where the wind fleet is likely to operate for some years to come.

**Figure 8. Comparison of model 1 predictions using 2014 real time data and a constant grid demand throughout the year of 34.34 $GW_e$**

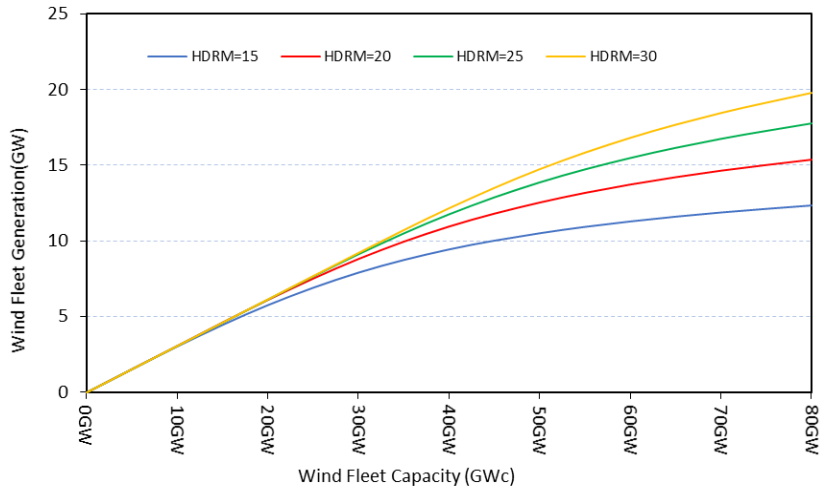

## 4. Model 2: Scaling Wind Generation Histograms

The finding of the previous section that replacing real time grid demand records with annual averages had little effect on the prediction of wind shedding leads to the possibility of a significantly simpler modelling approach which uses annual wind generation histograms as input rather than real time data. It is found that the simpler model produces only slightly less accurate results than the very much more complex real time model.

Table 2 shows the elements of a simple spreadsheet model which uses the annual wind generation histogram to calculate wind generation, $GW_e$, for a particular wind fleet capacity, $GW_c$, and Hdrm. In the example illustrated the wind generation histogram for 2014 appears in columns B and C, the entries in B being the centres of each generation range of 0.5 $GW_e$, and C the fraction of time during 2014 the wind fleet spent in each generation range. The elements in D represent the generation in each range.

The inputs to the spreadsheet are Hdrm (A4), wind fleet capacity for which we wish to calculate wind generation (A6), and wind fleet capacity seen by Gridwatch in 2014 (A9) (see Section 3). The wind multiple is A6/A9 (=A11), and a revised generation range B*A11 appears in column E. Since the wind fleet is restricted by Hdrm (A4) (20 in this example), each element of E is capped at this level, and the capped range appears in column F. The generation in each element of the revised range is F times the appropriate time fraction in column C, and appears in column G. Total generation, the sum of G1:G14, appears in G18, and is 15.132 $GW_e$ in this example.

Repeated use of the spreadsheet of Table 2 for a range of Hdrm (15 $GW_e$, 20 $GW_e$, 25 $GW_e$ and 30 $GW_e$) and of $GW_c$ (10 $GW_c$ to 80 $GW_c$) using the histograms of Figure 7 for the years 2013 to 2016 resulted in a set of $GW_e$ vs $GW_c$ curves for different values of Hdrm. Since the $GW_e$ vs $GW_c$ curves for the four years were similar, they were averaged and Figure 9 compares the four-year average real time results (curves without symbols) with the four-year averages of the simpler histogram model (curves with square symbols).

**Table 2. Illustration of spreadsheet to calculate generate wind generation (GW$_e$) for a GW$_c$=80 and Hdrm = 20 GW$_e$ using the 2014 wind generation histogram**

| | A | B | C | D | E | F | G |
|---|---|---|---|---|---|---|---|
| | | Centre of generation range(GW) | time fract in each range | Generation in range | Revised centre of generation range (GW) | Revised centre of generation range capped by Hdrm | Generation in new range GWe |
| 1 | Input data | 0.25 | 0.097819 | 0.024454856 | 2.522068096 | 2.522068096 | 0.246707244 |
| 2 | | 0.75 | 0.157028 | 0.117771327 | 7.566204288 | 7.566204288 | 1.188109223 |
| 3 | Headroom | 1.25 | 0.130231 | 0.16278886 | 12.61034048 | 12.61034048 | 1.642258356 |
| 4 | 20 | 1.75 | 0.103664 | 0.181411436 | 17.65447667 | 17.65447667 | 1.830127976 |
| 5 | GWc= | 2.25 | 0.087022 | 0.195799322 | 22.69861286 | 20 | 1.740438415 |
| 6 | 80 | 2.75 | 0.074165 | 0.203952537 | 27.74274905 | 20 | 1.483291178 |
| 7 | Wind fleet | 3.25 | 0.068589 | 0.222912834 | 32.78688525 | 20 | 1.371771288 |
| 8 | capacity | 3.75 | 0.063597 | 0.238488733 | 37.83102144 | 20 | 1.27193991 |
| 9 | 7.93 | 4.25 | 0.061805 | 0.262672932 | 42.87515763 | 20 | 1.236107918 |
| 10 | Wind multiple | 4.75 | 0.057015 | 0.270821357 | 47.91929382 | 20 | 1.140300452 |
| 11 | 10.08827238 | 5.25 | 0.050577 | 0.265527995 | 52.96343001 | 20 | 1.011535219 |
| 12 | | 5.75 | 0.035928 | 0.206584847 | 58.0075662 | 20 | 0.71855599 |
| 13 | | 6.25 | 0.011095 | 0.069340653 | 63.0517024 | 20 | 0.22189009 |
| 14 | | 6.75 | 0.001466 | 0.009894516 | 68.09583859 | 20 | 0.029317084 |
| 15 | | | | | | | |
| 16 | | | | | | | Total generation |
| 17 | | | | | | | at GWc=D80 |
| 18 | | | | | | | 15.13235034 |

**Figure 9. Comparison of the GW$_e$ vs GW$_c$ curves derived using the two different models**

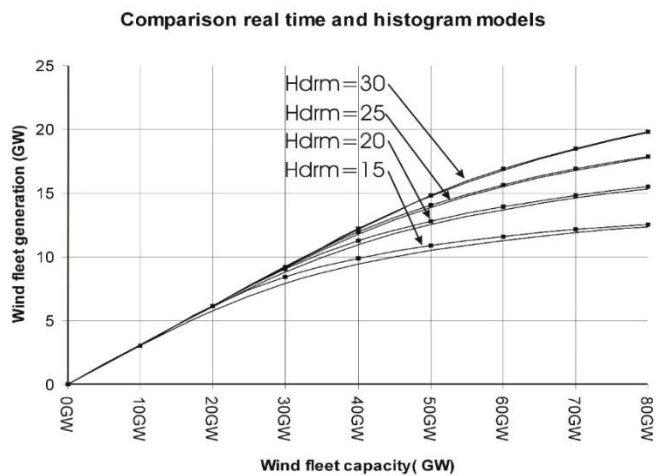

It may be seen in Figure 9 that the simpler histogram-based model predicts slightly less wind shedding than the real time models, particularly from small wind fleets, and Figure 3 enables us to understand why this should be so. At times of very low grid demand, such as on the night of 2$^{nd}$ Nov 2014, the real time-based model would have predicted the onset of wind shedding for a wind generation of only around 7 GW$_e$. By definition, the histogram-based models will not produce any wind shedding until the predicted wind generation exceeds Hdrm, which was about 20 GW$_e$ in 2014.

## 5. GW$_e$ vs GW$_c$ Curves and Wind Fleet Efficiency

The reason for using the models to produce the GW$_e$ vs GW$_c$ curves such as those seen in Figures 6 and 8 is that they enable us to quantify wind shedding and thereby provide a measure of wind fleet efficiency. The efficiency measure adopted in this paper is the incremental increase in wind generation produced by a unit increase in wind fleet capacity. This measure, which we call the wind fleet's Marginal Efficiency is by definition the gradient of the $GW_e$ vs $GW_c$ curve, from which it might be calculated directly. Figure 10 shows Marginal Efficiency curves derived from the real time $GW_e$ vs $GW_c$ curves of Figure 9.

**Figure 10. Marginal Efficiency curves for different values of Hdrm derived from the real time $GW_e$ vs $GW_c$ curves**

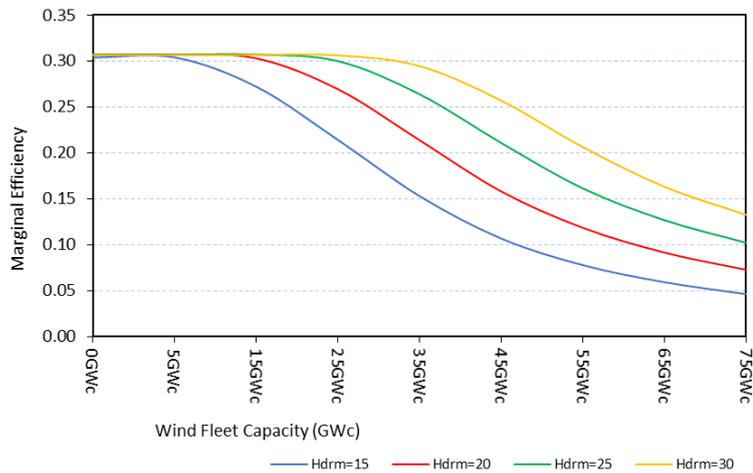

As we would expect, Figure 10 shows that the Marginal Efficiency of small wind fleets is equal to the wind fleet's load factor (0.3075), since there is no wind shedding from small wind fleets, but that the Marginal Efficiency decreases as the wind fleet increases in size. Of particular note is that the onset of Marginal Efficiency reductions is at much higher $GW_c$ values for larger values of Hdrm.

It is instructive, for illustrative purposes, to investigate the consequences of assuming a value for the lowest acceptable Marginal Efficiency, and Table 3 shows, in tabular form, the predictions of $GW_e$ and $GW_c$ should the lowest acceptable Marginal Efficiency be 0.2. Marginal Efficiency is a useful measure of how much of a wind fleet's potential output is available for use, and how much must be shed. Since the upper economic limit of the wind fleet is also likely to be determined by wind shedding, it is therefore likely to be closely related to Marginal Efficiency, a measure which our models enable us to predict.

**Table 3. Wind fleet parameters for a marginal efficiency of 0.2**

| Annual Average Headroom ($GW_e$) | $GW_c$ | $GW_e$ | $GW_e$/Hdrm |
|---|---|---|---|
| 15 | 27.22 | 7.33 | 0.4887 |
| 20 | 37.47 | 10.4 | 0.5197 |
| 25 | 47.17 | 13.3 | 0.532 |
| 30 | 56.57 | 16.09 | 0.532 |

The UK wind fleet's Hdrm is currently approximately 20 $GW_e$, which suggests that, for grid configurations as of today, the upper economic limit for the wind fleet will be reached when the wind fleet capacity is 37.47 $GW_c$, roughly twice the capacity of 17.7 $GW_c$ at the end of 2017. The resulting wind generation would then be 10.4 $GW_e$, close to the UK government's target set in 2007 of 10 $GW_e$ (Goodall, 2007). However, should the Hdrm increase to 25 GW, a possibility if much of the UK nuclear fleet reaches the end of its life without replacement during the 2020s, the upper economic capacity of the wind fleet will increase to 47.17 $GW_c$, and generation to 13.3 GW$_e$. The fourth column in Table 3, GW$_e$/Hdrm, provides an interesting rule of thumb that the upper economic generation of the UK wind fleet for a lowest Marginal Efficiency of 0.2 is approximately half the available Hdrm.

## 6. Competition between Wind and Solar Generation

In April 2017, the National Grid announced that an increase in solar generating capacity was leading to a significant reduction in grid demand. On occasions when the grid was unable to accommodate all wind and solar generation, wind farms would be paid not to generate (Gosden, 2017). 2017 was the first year for which solar generation was recorded by the Gridwatch website, these records providing for the first time a means of predicting how wind and solar generation are likely to interact in future.

In Figure 11, wind generation is shown stacked above base generation, with a new curve, grid demand + solar generation, seen above grid demand for a typical week in 2017. This provides an insight into just how much solar generation was already diminishing the wind fleet's available Headroom in 2017. Although the average annual solar generation was only 1.17 GW$_e$ in 2017, about a third of the average annual wind generation of 3.65 GW$_e$, peak wind and solar generations were frequently similar at about 8 GW$_e$. This explains why, as the wind and solar fleets increase in size, there will be an increasing number of occasions when their combined output will exceed grid demand, and wind generation will have to be curtailed. Figure 11 suggests that wind and solar generation will first compete for grid access at week-ends, when grid demand is lower than during the week.

**Figure 11. Contribution of wind and solar generation to grid demand during week 23 of 2017 (Monday 5$^{th}$ June to Sunday 11$^{th}$ June)**

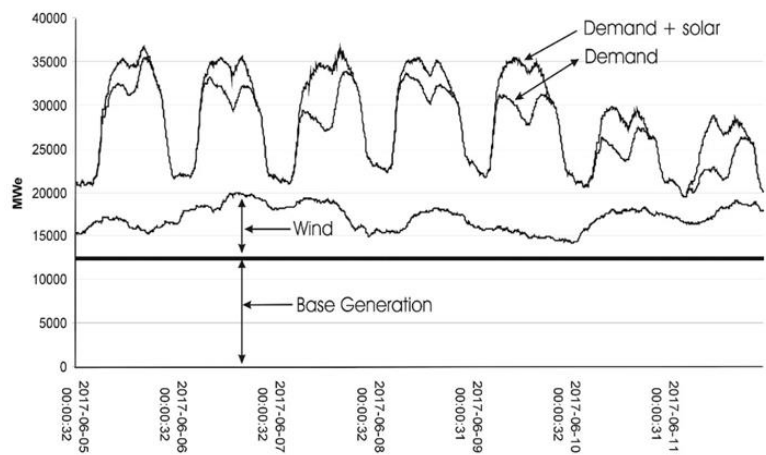

To investigate the interaction between wind and solar generation further, the real time model of Section 3 was modified to accept solar generation as an additional input, and wind generation was curtailed when wind plus solar generation exceeded grid demand. The weekly records for 2017 were downloaded from the Gridwatch website so that the annualised GW$_e$ vs GW$_c$ and Marginal Efficiency curves might be calculated for three different levels of solar generation viz

- no solar generation (no symbols in Figure 12)
- solar generation as in 2017 (square symbols in Figure 12)
- twice the solar generation of 2017 (triangular symbols in Figure 12).

The resulting $GW_e$ vs $GW_c$ and Marginal Efficiency curves for different values of Hdrm are shown in Figure 12. If, as in the previous worked example, the lowest acceptable Marginal Efficiency of the wind fleet was deemed to be 0.2, Figure 12 suggests that a doubling of solar generation, an increase in average annual generation of 1.17 $GW_e$, will reduce the upper economic wind fleet capacity by about 2 $GW_c$, and reduce wind generation by about 0.4 $GW_e$. Clearly solar generation significantly reduces the Hrdm available to the wind fleet, and future investments in the wind and solar fleets will need to be coordinated to avoid wasteful competition between these two high cost sources of generation.

In 2017, the wind fleet capacity was 17.7 $GW_e$ and Hdrm approximately 20 $GW_e$. Figure 12 (right) suggests we must therefore expect a gradual increase in the amount of wind shedding in future as the wind and solar fleets increases in size. Examples of wind and solar generation coming close to using all the available Headroom may be seen in Figure 11 during week 23 of 2017. On the night of 6th June 2017, it was wind generation alone which came close to exceeding the available Headroom, while during daylight hours on 11th June it was the combination of high wind generation, high solar generation and low grid demand which caused the reduction of available Headroom.

**Figure 12. $GW_e$ vs $GW_c$ and Marginal Efficiency curves for different values of Hdrm and different levels of solar generation**

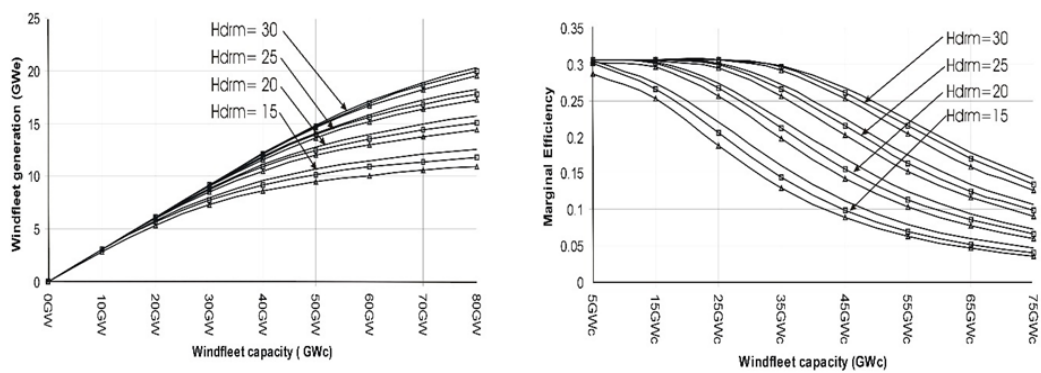

Source: Gridwatch

## 7. The Wind Fleet's Potential to Reduce UK Carbon Dioxide Emissions

The purpose of the UK wind fleet is to decarbonise the electricity grid, and the models described earlier provide a means of prediction the wind fleet's future effectiveness in achieving this objective. Table 4 shows the progress between 1990 and 2017 in reducing carbon dioxide emissions for the three main contributory sources viz Electricity Generation, Business and Transport.

**Table 4. Sources of UK carbon dioxide emissions in 1990 and 2017 in MT per annum**

|  | 1990 | 2017 | Difference |
|---|---|---|---|
| Electricity Generation | 203 | 71.8 | (-121.2) |
| Business | 111.9 | 65.8 | (-46.1) |
| Transport | 125.3 | 124.4 | (-0.9) |
| **Total** | **594.1** | **366.9** | **(-227.2)** |

Source: National Statistics UK (2018)

A recent review (Papaioannou et al., 2017) found different authorities making a variety of assumptions about the emissions resulting from a GWe of coal and gas generation, as shown in Table 5. In the following analysis we shall use the column 1 values in Table 5 since they produce the closest fit between the UK government emissions figure of 71.8 MT per annum. of Table 4 and the 76.15 MT per annum. inferred from UK government records for electricity generation, as shown in Table 6.

**Table 5. Range of estimates in UK carbon dioxide emissions (MT per annum/$GW_e$) from coal fired and gas fired generation**

|  | Minimum | Mean | Max |
|---|---|---|---|
| Coal fired generation | 7.88 | 8.44 | 8.99 |
| Gas fired generation | 3.67 | 4.27 | 4.87 |

Source: Papaioannou et al. (2017)

Assuming that the 5.66 $GW_e$ of wind generation in 2017 displaced coal rather than gas generation, the 17.7 $GW_c$ wind fleet in 2017 would have been responsible for a reduction in carbon dioxide emissions of 5.66 * 7.88= 44.6 MT. The wind fleet's Headroom in 2017 in Table 6 was 23.46 $GW_e$, being made up of coal=2.58 $GW_e$, gas=15.21 $GW_e$ and wind = 5.66 $GW_e$, but a number of factors make it impossible to predict what the Headroom will be in future decades. Changes in grid demand, which has been falling steadily in recent years, suggest that Headroom might be as low as 20 $GW_e$ by 2020, but Headroom is then likely to rise during the 2020s due to the retirement of nuclear reactors without replacement and the replacement of petrol and diesel cars by electric vehicles, provided the additional grid demand is met by gas fired generation (which would increase Headroom).

**Table 6. Average generation by sector in 2017 and carbon dioxide emissions calculated using column I figures in Table 5**

| Source of generation | $GW_e$ | Carbon Dioxide Emission (MT) |
|---|---|---|
| Coal | 2.58 | 20.33 |
| Nuclear | 8.025 |  |
| Gas | 15.21 | 55.82 |
| Wind | 5.66 |  |
| Other renewable | 5.62 |  |
| Other | 1.31 |  |
| **Total** | **38.34** | **76.15** |

Source: Department for Business et al. (2018)

In view of the uncertainties about the level of Headroom in future years, the models were used to predict the impact on carbon dioxide emissions of a range of Headrooms from 15$GW_e$ to 30$GW_e$, which covers all likely expectations in decades to come.

The reference point for these calculations is the wind fleet's condition in 2017 ( = 23.45$GW_e$, $GW_c$= 17.7 and MT of emissions = 76.1 MT per annum) and is represented by point A0 in Figure 13. The reference points for the other Headroom curves are calculated from point A0 by adjusting the level of gas generation. Thus, A1, the reference point for Headroom = 30, has additional gas generation of 30- 23.45= 6.55 $GW_e$. This gives an additional 6.55*3.67 = 24.03 of emissions and a total emission of 104.13 MT per annum at point A1. The starting points of the other Headrooms are calculated in a similar manner to give

- A1 (Hdrm=30, $GW_c$=17.7, MT= 104.1)
- A2 (Hdrm =25, $GW_c$=17.7, MT=81.79)
- A3 (Hdrm=20, $GW_c$= 17.7, MT=63.44)
- A4 (Hdrm=15, $GW_c$=17.7, MT= 45.09).

For each value of Hdrm, the model calculates emissions for different levels of wind fleet capacity, $GW_c$. Initially, the increase in wind generation replaces coal generation at 7.88 MT per annum/$GW_e$, until the 2.58 $GW_e$ of coal generation of 2017 has been reduced to zero, represented by points B1 to B4 in Figure 13.

- B1 (Hdrm=30, $GW_c$=27.31, MT= 79.80)
- B2 (Hdrm =25, $GW_c$=27.43, MT=61.49)
- B3 (Hdrm=20, $GW_c$=28.38, MT=43.14)
- B4 (Hdrm =15, $GW_c$=32.61, MT= 24.791).

As the wind fleet increases further in size it replaces gas generation at 3.67 MT per annum/$GW_e$, explaining the reduction in slope of the curves to the right of B1 to B4 and the decrease in effectiveness of the wind fleet in reducing emissions (see Table 7).

**Table 7. The wind fleet's efficiency in reducing emissions (MT pa/ $GW_c$)**

|  | Hdrm=15 | Hdrm=20 | Hdrm=25 | Hdrm=30 |
|---|---|---|---|---|
| Points A1-A4 to B1-B4 (i.e. wind displacing coal fired generation) | 1.36 | 1.9 | 2.08 | 1.98 |
| Between 30$GW_c$ and 40 $GW_c$ | 0.56 | 0.78 | 0.957 | 1.07 |
| Between 40 $GW_c$ and 50$GW_c$ | 0.38 | 0.57 | 0.763 | 0.928 |
| Between 50$GW_c$ and 60$GW_c$ | 0.286 | 0.42 | 0.579 | 0.743 |
| Between 60$GW_c$ and 70 $GW_c$ | 0.21 | 0.32 | 0.458 | 0.584 |
| Between 70$GW_c$ and 80$GW_c$ | 0.17 | 0.26 | 0.366 | 0.481 |

Although it is beyond the scope of this paper to take a view on what will be considered the lowest economic efficiency of the wind fleet in future, it is instructive to consider the consequences of, for example, 0.75 MT pa/$GW_c$ being considered the lowest economic efficiency. For a Hdrm of 20, perhaps the most likely value in the early 2020s, this would give an upper economic limit of the wind fleet of 35 $GW_c$. At this point, point C1 in Figure 13, wind generation would be 9.735 $GW_e$ and the reduction in carbon dioxide emissions 37 MT/pa.

The points along the line C1 to C2 predict the upper economic limits of the wind fleet should the Hdrm increase either due to retirement of nuclear reactors without replacement or the large-scale replacement of petrol and diesel cars by electric vehicles (EVs). It is looking increasingly unlikely that new nuclear reactors will be brought on line before the ageing Advanced Cooled Reactors have to be retired from service in the late 2020s, and it is quite feasible therefore to consider an increase in Hdrm to 25, and a consequent increase in the upper economic limit of the wind fleet to 45$GW_c$.

Although EVs are being introduced only very slowly in the UK it is possible that their widespread deployment in future could increase Hdrm to 30 $GW_c$ when the maximum economic deployment of the wind fleet would be 55$GW_c$ (C2 in Figure 13). In order to calculate the consequent reductions in carbon dioxide emissions we need to consider not only the direct reductions in emissions from the transport fleet but also the increase in emissions from an additional generation needed to power the EVs.

**Figure 13. Carbon dioxide emissions for Hdrm's of 15, 20, 25 and 30 GW and a range of wind fleet capacities up to 80GW$_c$**

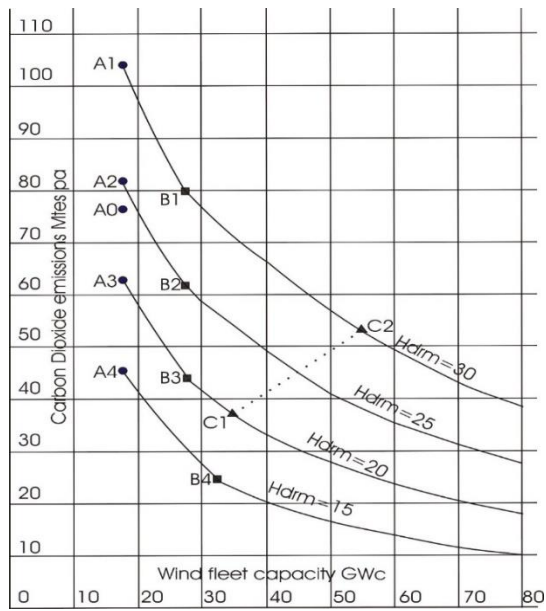

According MacKay (2009), an EV driven under average UK conditions requires around 10 kWh a day and, should this be the case, an additional generation of 10 GW$_e$ would be sufficient to power 24 million EVs. In 2016, there were 25.8 million petrol and diesel cars on the UK roads (BBC News, 2016) so 10 GW$_e$ should be sufficient to power nearly all the replacement EVs. At point C2, the wind fleet would be generating 15.5 GW$_e$, 5.81 GW$_e$ more than the 9.69 GW$_e$ at point C1. This means that of the 10 GW$_e$ needed to power 24M EVs, 5.81 GW$_e$ would be from additional wind generation and only 4.19 GW$_e$ would be needed from additional gas generation. At 3.67 MT per annum/GW$_e$ the additional 4.19 GW$_e$ of gas generation would lead to an increase of 4.19*3.67= 15.4 MT per annum in emissions.

The 25.8 million petrol and diesel cars in the UK generated 68.5 MT of carbon dioxide in 2016 (Society of Motor Manufacturers and Traders, 2018), so the conversion of 24 petrol and diesel cars to EVs should lead to a direct reduction of 63.7 MT per annum in transport emissions. Offsetting the additional 15.37 MT per annum from electricity generation gives a net reduction in emissions of 48.3 MT per annum in moving from point C1 to C2 on Figure 13.

Given the slow growth in introduction of EVs worldwide, it is likely to be some years before sizeable numbers of EVs will be seen on UK roads. However, the dotted line C1 to C2 in Figure 13 is useful in showing the trajectory which would be expect to be followed as more EVs are introduced, with a net emission reduction in carbon dioxide emissions of 2.01 MT per annum/million cars replaced by EVs.

## 8. Concluding Remarks

Two different models, each applied to the real time records for the years 2013 to 2016, reveal that the records for only a single year are required to develop a dataset of general applicability. An explanation for this perhaps surprising empirical finding is that, although wind in the UK appears to be random from week to week and month to month, annual distributions change very little between different bands of wind generation. This finding also leads to a much simpler modelling approach which uses wind generation histograms rather than real time data as model inputs. Despite the modelling approaches being so radically different, they give similar predictions about the performance of the wind fleet as it increases in size.

It has been suggested that the upper economic limit of the UK wind fleet is a function of wind fleet capacity (Judge, 2016) or wind penetration (Korchinski, 2013), but the models suggest that such formulations are over simplistic. The upper economic limit of the wind fleet will depend on the Headroom available to the wind fleet, which will in turn depend on a number of factors that are impossible to predict with any certainty. Because of the uncertainty about the future level of Headroom, the paper presents results for different values of Headroom ranging from 15 $GW_e$ to 30 $GW_e$, which should cover all likely eventualities in the decades to come.

It would appear likely that the Headroom in the early 2020s will be around 20 $GW_e$ and, should this be the case, the models predict that the upper economic limit of the wind fleet will be about 35 $GW_e$, around twice its size in 2017. Indeed, if all the turbines either in service or having been given consent at the end of 2017 come into service, they will bring the wind fleet up to its upper economic limit. Any further investment will lead to a significant decrease in efficiency of the wind fleet because of the increasing need to shed wind generation when it is surplus to the requirements of the grid unless other factors cause the wind fleet's Headroom to increase.

What would increase the wind fleet's Headroom, and therefore its upper economic limit, would be either the retirement of the UK's ageing nuclear reactors without replacement, or the introduction of a large number of EVs (to replace diesel and petrol cars). The models suggest an increase of 5 $GW_e$ in Headroom would increase the upper economic limit of the wind fleet by 10 $GW_c$ (to 45 $GW_c$) and an increase of 10 $GW_e$ in Headroom would increase the upper economic limit of the wind fleet by 20 $GW_c$ (to 55 $GW_c$).

Also discussed is the reduction in efficiency of the wind fleet in reducing carbon dioxide emissions as the wind fleet increases in size. This is a consequence of the wind fleet first displacing coal generation but then replacing gas generation once coal generation has been completely displaced.

A relatively recent occurrence has been the competition between the wind and solar fleets for access to the grid (Gosden, 2017). The 2017 real time records, the first to include solar generation, show that the solar fleet is already beginning to reduce the Headroom available to the wind fleet. It is clear that future investment in these two sources of renewable energy will need to be coordinated if one is not to damage the economics of the other.

**Abbreviations**

| | |
|---|---|
| CCGT | Combined Cycle Gas Turbine |
| Hdrm | The wind fleet's (average annual) Headroom |
| $GW_c$ | Wind fleet capacity in Gigawatts |
| $GW_e$ | Wind generation in Gigawatts |
| OCGT | Open Cycle Gas Turbine |

**Terms with Special or Restricted Meanings**

| | |
|---|---|
| Base generation | Summation of generation from sources given preferential access to the grid e.g. nuclear generation |
| Grid demand | The demand on the grid as recorded by Gridwatch |
| Gridwatch | The website where UK generation records at 5-minute intervals may be accessed |
| Headroom | The difference between grid demand and base generation |
| Load factor | The wind fleet's annual average generation divided by its capacity |
| Marginal Efficiency | Incremental increase in wind generation for an incremental increase in wind fleet capacity, and the gradient of the $GW_e$ vs $GW_c$ curves from which the Marginal Efficiency may be calculated directly |
| Wind generation | Wind fleet generation as recorded by the Gridwatch website (roughly 60% of total UK wind generation) |

**References**

Barnhart, C. J., Dale, M., Brandt, A. R. & Benson, S. M. 2013. The energetic implications of curtailing versus storing solar-and wind-generated electricity. *Energy & Environmental Science,* 6(10)**,** pp 2804-2810.

BBC News. 2016. *Cars on England's roads increase by almost 600,000 in a year* [Online]. London: BBC. Available: https://www.bbc.com/news/uk-england-35312562 [Accessed 17 June 2018].

Department for Business, Energy & Industrial Strategy. 2018. *UK Energy statistics; statistical press release – March 2018* [Online]. London: Department for Business, Energy & Industrial Strategy,. Available: https://www.gov.uk/government/news/uk-energy-statistics-statistical-press-release-march-2018 [Accessed 17 June 2018].

Edwards, V. 2017. Elon Musk's Tesla to build giant battery for SA. *The Australian*, 7 July 2017, Available: http://www.theaustralian.com.au/business/mining-energy/elon-musks-tesla-to-help-build-worlds-largest-lithium-ion-battery-for-sa/news-story/ff818dd5da6d8ebb1bf880d7900f1672 [Accessed 11 October 2017].

Eller, A. & Gauntlett, D. 2017. Energy Storage Trends and Opportunities for Emerging Markets. IFC Conference Edition, Navigant Consulting (Boulder, USA).

Gosden, E. 2017. Wind farms could be paid to stop producing power. *The Times* 7 April 2017, Available: [https://www.thetimes.co.uk/article/wind-farms-could-be-paid-to-stop-producing-power-qqm5dl7wk](https://www.thetimes.co.uk/article/wind-farms-could-be-paid-to-stop-producing-power-qqm5dl7wk) [Accessed 12 October 2017].

Gridwatch. 2017. *G.B. National Grid Status (data courtesy of Elexon portal and Sheffield University)* [Online]. Templar. Available: [http://www.gridwatch.templar.co.uk/](http://www.gridwatch.templar.co.uk/) [Accessed 4 April 2017].

Judge, B. 2016. *When it comes to nuclear's future, it's time to think small* [Online]. London: The Times. Available: [http://www.thetimes.co.uk/article/when-it-comes-to-nuclears-future-its-time-to-think-small-ntnqhvnrr](http://www.thetimes.co.uk/article/when-it-comes-to-nuclears-future-its-time-to-think-small-ntnqhvnrr) [Accessed 17 June 2018].

Korchinski, W. 2013. The Limits of Wind Power. Policy Study 403, Adam Smith Institute (Washington).

MacKay, D. 2009. *Sustainable Energy - Without the Hot Air*, Cambridge: UIT Cambridge.

Marcacci, S. 2013. *Study: Battery Energy storage benefits soar not wind* [Online]. Clean Technica. Available: [https://cleantechnica.com/2013/09/13/study-battery-energy-storage-works-for-solar-but-not-wind/](https://cleantechnica.com/2013/09/13/study-battery-energy-storage-works-for-solar-but-not-wind/) [Accessed 17 June 2018].

National Statistics UK. 2018. 2017 UK greenhouse gas emissions; provisional figures. Department for Business, Energy & Industrial Strategy (London).

Papaioannou, V., Coker, P., Potter, B. & Livina, V. 2017. Time-varying grid carbon intensity of the UK for the years 2009-2016. *SEM 2017.*

Royal Academy of Engineering. 2014. Wind Energy Report. Royal Academy of Engineering (London).

Society of Motor Manufacturers and Traders. 2018. New car $CO_2$ report 2018: the 17th report. SMMT (London).

Stephens, A. D. & Walwyn, D. 2017. The Security of the United Kingdom's Electricity Imports under Conditions of High European Demand. *arXiv,* 1802.07457**,** pp 1-8.

Stephens, A. D. & Walwyn, D. R. 2016. Wind Energy in the United Kingdom: Modelling the Effect of Increases in Installed Capacity on Generation Efficiency. *arXiv,* 1611.04174**,** pp 1-12.